\documentclass[a4paper,12pt]{article}

\usepackage{geometry}
\usepackage{amsfonts}
\usepackage{amsmath,amssymb}
\usepackage{amsthm}
\usepackage{authblk}
\usepackage{hyperref}
\usepackage{bm}
\usepackage{cases}
\usepackage{cite}
\usepackage{mathrsfs}
\usepackage[all]{xy}

\allowdisplaybreaks

\theoremstyle{plain}
\newtheorem{prop}{Proposition}
\newtheorem{coro}{Corollary}

\newtheorem{lemma}{Lemma}

\theoremstyle{remark}
\newtheorem{remark}{Remark}

\title{Nonlocal Symmetries of Two 2-component Equations of Camassa-Holm Type}

\author{Ziqi Li}
\author{Kai Tian\thanks{Corresponding author (Email: \texttt{tiankai@lsec.cc.ac.cn})}}

\affil{\small Department of Mathematics, China University of Mining and Technology, Beijing, 100083, \par People's Republic of China}

\date{February, 2024}

\begin{document}
\maketitle

\begin{abstract}
	For a 2-component Camassa-Holm equation, as well as a 2-component generalization of the modified Camassa-Holm equation, nonlocal infinitesimal symmetries quadratically depending on eigenfunctions of linear spectral problems are constructed from functional gradients of spectral parameters. With appropriate pseudo-potentials, these nonlocal infinitesimal symmetries are prolonged to enlarged systems, and then explicitly integrated to generate symmetry transformations in finite form for enlarged systems. As implementations of these finite symmetry transformations, some kinds of nontrivial solutions and B\"{a}cklund transformations are derived for both equations.

	\vspace{8pt}
	{\footnotesize
	
	}
	
\end{abstract}

\section{Introduction}
Integrable equations of Camassa-Holm (CH) type have been studied extensively in the past three decades, and shown to admit remarkable features in various aspects. The CH equation itself describes the unidirectional propagation of shallow water waves over a flat bottom \cite{camaholm93}, and also arises as an equation of the geodesic flow for the $H^1$ right-invariant metrics on the Bott–Virasoro group \cite{misio98,kour99}. Its integrability was explained in a geometrical viewpoint based on the $\mathfrak{sl}(2)$-valued linear spectral problem \cite{reyes02}. Such equations possess infinitely many local and/or nonlocal (infinitesimal) symmetries, which bring us with effective tools to reveal sophisticated properties of such equations. The CH equation, as well as the modified CH equation, was shown to admit nonlocal infinitesimal symmetries depending on pseudo-potentials which are closely related to linear spectral problems \cite{reyes07,bies12,lou22}. Moreover, symmetry transformations in finite form were established by integrating these nonlocal infinitesimal symmetries, and may lead to nontrivial solutions, B\"{a}cklund/Darboux transformations for both equations \cite{reyes09,litian22}. In this work, we manage to discuss two 2-component equations of CH type by an approach via nonlocal symmetries. Eigenfunctions of linear spectral problems would play roles of pseudo-potentials \cite{galas92}, and nonlocal infinitesimal symmetries defined in terms of them will be constructed. 

A 2-component CH (2-CH) equation we will consider is formulated as
\begin{equation} \label{eq:2ch}
	\left\{	\begin{aligned}
		m_{t} =&\; -um_{x}-2u_{x}m+\sigma \rho\rho_{x} , \quad m = u - u_{xx}, \\
		\rho_{t} =&\; -(\rho u)_{x} ,
	\end{aligned} \right. 
\end{equation}
where $u = u(x,t)$, $\rho = \rho(x,t)$, and the parameter $\sigma = \pm 1$, while subscripts stand for partial derivatives with respect to the indicated independent variables. It was claimed that equation \eqref{eq:2ch} stems from the shallow water theory, and has a linear spectral problem, as well as a bi-Hamiltonian structure \cite{consivan08}. If $\rho = 0$, then equation \eqref{eq:2ch} is reduced to the CH equation (with $\kappa =0 $), i.e.
\begin{equation}\label{eq:CH}
	m_{t} = -um_{x} - 2u_{x}m,\quad m = u - u_{xx}. 
\end{equation}
Equation \eqref{eq:2ch} with $\sigma = - 1$ was originally proposed as a tri-Hamiltonian dual system to the Ito equation \cite{olvros96}, while the case of $\sigma = 1$ was constructed in the deformation theory of bi-Hamiltonian structure of hydrodynamic type \cite{chenliuzhang06}, and also obtained by the bi-Hamiltonian approach \cite{falqui06}. Some kinds of solutions were constructed for equation \eqref{eq:2ch} (setting $\sigma$ to be either $1$ or $-1$) through various approaches, such as the method via the reciprocal link between equation \eqref{eq:2ch} with $\sigma=1$ and the first negative flow of the AKNS hierarchy \cite{chenliuzhang06,wu06}, the inverse scattering transformation \cite{holmivan11}, bilinear method \cite{matsuno17}, Darboux transformation  \cite{wangliliu20}, B\"{a}cklund transformation \cite{wangThesis21,wang23} etc.. 

The other equation we will discuss is a 2-component generalization of the modified CH equation, given by \cite{tianliu13}
\begin{equation}\label{eq:2mch}
	\left\{	\begin{aligned}
		m_{t} =&\; (mqr)_{x}, &&  m = q - q_{x} ,\\
		n_{t} =&\; (nqr)_{x}, &&  n = - r - r_{x} ,
	\end{aligned} \right.
\end{equation}
where $q = q(x,t)$ and $r = r(x,t)$. Let $q = u_{x} + u$ and $r = - u_{x} + u$, then $n = -m = - (u - u_{xx})$, and equation \eqref{eq:2mch} is reduced to the modified CH equaiton \cite{olvros96,qiao06}, namely
\begin{equation}\label{eq:mch}
	m_{t} = [m(u_{x}^2 - u^2)]_{x}, \quad  m = u - u_{xx}.
\end{equation}
In the framework of tri-Hamiltonian duality, equation \eqref{eq:2mch} was obtained as a dual system to the Wadati-Konno-Ichikawa hierarchy \cite{tianliu13}. If we introduce $q = u_{x} + u$ and $r = v_{x} - v$, then equation \eqref{eq:2mch} is rewritten as 
\begin{equation}\label{eq:sqq}
	\left\{	\begin{aligned}
		m_{t} =&\; [m(u_{x}v_{x} - u_{x}v + uv_{x} - uv)]_{x}, &&  m = u - u_{xx} ,\\
		n_{t} =&\; [n(u_{x}v_{x} - u_{x}v + uv_{x} - uv)]_{x}, &&  n = v - v_{xx} ,
	\end{aligned} \right.
\end{equation}
which was first introduced by Song, Qu and Qiao, and shown to be integrable by presenting a linear spectral problem and an infinite set of conserved quantities \cite{sqq11}. Equation \eqref{eq:sqq} was shown to have interlacing peakons by analysing its spectral and inverse spectral problems \cite{chang16}.

The 2-CH equation \eqref{eq:2ch} is an isospectral flow yielded from its linear spectral problem, so the spectral parameter qualifies as a conserved quantity (functional). There is the same claim for equation \eqref{eq:2mch}. Given any Hamiltonian system, the Hamiltonian operator would cast functional gradients of conserved quantities into its infinitesimal symmetries. This mechanism is also applicable to spectral parameters, and would produce nonlocal infinitesimal symmetries defined in terms of eigenfunctions of linear spectral problems. By this approach, nonlocal infinitesimal symmetries quadratically depending on eigenfunctions of linear spectral problems were constructed for some super integrable equations \cite{zhoutianli22}. For backgrounds on nonlocal symmetries of differential equations, one may refer to \cite{KrasilVinog89} and references therein.
 
The paper is organized as follows. In section \ref{sec:2}, for the 2-CH equation \eqref{eq:2ch} and equation \eqref{eq:2mch} respectively, functional gradients of spectral parameters are calculated, and mapped by Hamiltonian operators to nonlocal infinitesimal symmetries. Furthermore, these symmetries are reduced to some quadratically depending on eigenfunctions of linear spectral problems. In section \ref{sec:3}, by introducting appropriate pseudo-potentials in each case, nonlocal infinitesimal symmetries are prolonged to linear spectral problems and equations governing pseudo-potentials, and then explicitly integrated to produce symmetry transformations in finite form for enlarged systems. By use of these finite symmetry transformations, we obtain some nontrivial solutions from trivial ones. In section \ref{sec:4}, B\"{a}cklund transformations are derived for the 2-CH equation \eqref{eq:2ch} and equation \eqref{eq:2mch} from finite symmetry transformations obtained in section \ref{sec:3}. Conclusions and discussions are given in section \ref{sec:5}.

\section{Nonlocal infinitesimal symmetries} \label{sec:2}
On the basis of bi-Hamiltonian structures and linear spectral problems, nonlocal infinitesimal symmetries are constructed in detail for the 2-CH equation \eqref{eq:2ch}. Parallel results are presented briefly for equation \eqref{eq:2mch}.

\subsection{2-CH equation \eqref{eq:2ch}}\label{subsec:2:1}
The 2-CH equation \eqref{eq:2ch} is formulated as Hamiltonian systems in two different ways \cite{olvros96,chenliuzhang06,falqui06,consivan08}
\begin{equation*}
	\begin{pmatrix}
		m_{t} \\ \rho_{t}
	\end{pmatrix} = \mathcal{B}_{1} \begin{pmatrix}
		\delta_{m} \\ \delta_{\rho} 
	\end{pmatrix} \frac{1}{2}\int u\left(-u_{x}^{2} - u^2 + \sigma \rho^{2}\right) \mathrm{d}x = \mathcal{B}_{2} \begin{pmatrix}
		\delta_{m} \\ \delta_{\rho} 
	\end{pmatrix} \frac{1}{2}\int\left(2mu - \sigma \rho^{2} \right)\mathrm{d}x,
\end{equation*}
where a pair of compatible Hamiltonian operators $\mathcal{B}_{1}$ and $\mathcal{B}_{2}$ are given by
\begin{equation}\label{hamop:2ch}
	\mathcal{B}_{1}= \begin{pmatrix}
		\partial_{x}-\partial^{3}_{x} & 0 \\
		0 & -\sigma \partial_{x}	
	\end{pmatrix}, \quad \mathcal{B}_{2} = \begin{pmatrix}
		-m\partial_{x}-\partial_{x}m & - \rho \partial_{x} \\
		-\partial_{x}\rho & 0	
	\end{pmatrix}.
\end{equation}
It has the linear spectral problem (with the spectral parameter $\lambda$) \cite{schiff96,chenliuzhang06,consivan08}
\begin{subequations}  
	\begin{align}
		\begin{pmatrix}
			f \\g	
		\end{pmatrix}_{x} =&\; \mathbb{L} \begin{pmatrix}
			f \\g	
		\end{pmatrix},\quad  \mathbb{L} \equiv \begin{pmatrix}
			0 & 1 \\
			\frac{1}{4}-\lambda m+\lambda^{2}\sigma\rho^{2} & 0
		\end{pmatrix}, \label{sp:2ch:a}\\
		\begin{pmatrix}
			f \\g	
		\end{pmatrix}_{t} =&\; \mathbb{M} \begin{pmatrix}
			f \\g	
		\end{pmatrix},\quad \mathbb{M} \equiv \begin{pmatrix}
			\frac{u_{x}}{2} & - \frac{1}{2\lambda} - u\\
			- \frac{1}{8\lambda} + u\left(\frac{1}{4} + \lambda m - \lambda^2\sigma\rho^2\right) - \frac{\lambda}{2}\sigma\rho^{2} &- \frac{u_{x}}{2}
		\end{pmatrix}. \label{sp:2ch:b}
	\end{align}
\end{subequations}
Under the isospectral assumption that the spectral parameter $\lambda$ is independent of $x$ and $t$, equations \eqref{sp:2ch:a} and \eqref{sp:2ch:b} are compatible, i.e. the zero-curvature equation
\begin{equation}\label{zc:2ch}
	\mathbb{L}_{t}-\mathbb{M}_{x}+[\mathbb{L},\mathbb{M}]=0
\end{equation}
holds identically for arbitrary $\lambda$, if and only if $(m,\rho)$ solves the 2-CH equation \eqref{eq:2ch}. To calculate the functional gradient of $\lambda$, we also need the adjoint spectral problem
\begin{subequations}  
	\begin{align}
		\begin{pmatrix}
			\hat{f} & \hat{g}	
		\end{pmatrix}_x =&\; - \begin{pmatrix}
			\hat{f} & \hat{g}	
		\end{pmatrix}\mathbb{L}, \label{adsp:2ch:a}\\
		\begin{pmatrix}
			\hat{f} & \hat{g}	
		\end{pmatrix}_t =&\; - \begin{pmatrix}
			\hat{f} & \hat{g}	
		\end{pmatrix}\mathbb{M}. \label{adsp:2ch:b}
	\end{align}
\end{subequations}
The compatible conditon of \eqref{adsp:2ch:a} and \eqref{adsp:2ch:b} is still the zero-curvature equation \eqref{zc:2ch}. Between the linear spectral problem \eqref{sp:2ch:a}\eqref{sp:2ch:b} and the adjoint one \eqref{adsp:2ch:a}\eqref{adsp:2ch:b}, there is an interesting connection stated as follows.
\begin{lemma}\label{lem:1}
	If $(f,g)^{\top}$ solves the spectral problem \eqref{sp:2ch:a}\eqref{sp:2ch:b}, then $(-g,f)$ fulfils the adjoint spectral problem \eqref{adsp:2ch:a}\eqref{adsp:2ch:b}.
\end{lemma}

Let's first calculate the variational derivative of $\lambda$ with respect to $m$. Given an arbitrary small perturbation $m + \varepsilon \Delta m$, then from \eqref{sp:2ch:a} we derive
\begin{align}
	\begin{pmatrix}
		f_{\ast}[\Delta m] \\ g_{\ast}[\Delta m]
	\end{pmatrix}_{x} =&\; \begin{pmatrix}
		0 & 1 \\
		\frac{1}{4}-\lambda m+\lambda^{2}\sigma \rho^{2} & 0
	\end{pmatrix} 	\begin{pmatrix}
	f_{\ast}[\Delta m] \\ g_{\ast}[\Delta m]
	\end{pmatrix} \notag\\
	&\;\qquad + \begin{pmatrix}
		0 & 0 \\
		-\lambda \Delta m - \left(m - 2\lambda\sigma \rho^{2}\right)\langle \Delta m,\delta_{m}\lambda\rangle & 0
	\end{pmatrix}  \begin{pmatrix}
		f \\ g 
		\end{pmatrix}, \label{dd:2ch}
\end{align}	
where $f_{\ast}[\Delta m]$ ($g_{\ast}[\Delta m]$ respectively) stands for the directional derivative of $f$ ($g$ respectively), and the pairing $\langle \Delta m,\delta_{m}\lambda\rangle$ is defined as 
\begin{equation*}
	\langle \Delta m,\delta_{m}\lambda\rangle \equiv \int \Delta m(\delta_{m}\lambda)\mathrm{d}x.
\end{equation*}
Left-multiplying both sides of \eqref{dd:2ch} by $(\hat{f},\hat{g})$ and integrating with respect to $x$, we have
\begin{align}
	&\; \int \begin{pmatrix}
		\hat{f} & \hat{g}	
	\end{pmatrix}\begin{pmatrix}
		f_{\ast}[\Delta m] \\ g_{\ast}[\Delta m]
	\end{pmatrix}_{x}\mathrm{d}x = \int \begin{pmatrix}
	\hat{f} & \hat{g}	
	\end{pmatrix}\begin{pmatrix}
		0 & 1 \\
		\frac{1}{4}-\lambda m+\lambda^{2}\sigma \rho^{2} & 0
	\end{pmatrix} 	\begin{pmatrix}
		f_{\ast}[\Delta m] \\ g_{\ast}[\Delta m]
	\end{pmatrix} \mathrm{d}x \notag\\
	&\;\qquad\qquad\qquad\qquad + \int \begin{pmatrix}
		\hat{f} & \hat{g}	
	\end{pmatrix} \begin{pmatrix}
		0 & 0 \\
		-\lambda \Delta m - \left(m - 2\lambda\sigma \rho^{2}\right)\langle \Delta m,\delta_{m}\lambda\rangle & 0
	\end{pmatrix}  \begin{pmatrix}
		f \\ g 
	\end{pmatrix} \mathrm{d}x . \label{ff:2ch}
\end{align}
Integrating the left hand side of \eqref{ff:2ch} by parts with vanishing boundary conditions, and taking \eqref{adsp:2ch:a} into account, we obtain
\begin{equation*}
	\int \begin{pmatrix}
		\hat{f} & \hat{g}	
	\end{pmatrix} \begin{pmatrix}
	0 & 0 \\
	-\lambda \Delta m - \left(m - 2\lambda\sigma \rho^{2}\right)\langle \Delta m,\delta_{m}\lambda\rangle & 0
	\end{pmatrix}  \begin{pmatrix}
	f \\ g 
	\end{pmatrix} \mathrm{d}x=0,
\end{equation*}
or equivalently
\begin{equation}\label{vd:2ch}
	- \langle \Delta m,\delta_{m}\lambda\rangle \int (m - 2\lambda\sigma \rho^{2})f\hat{g} \mathrm{d}x-\int \lambda \Delta m f\hat{g} dx = 0.
\end{equation}
Since $\Delta m$ is arbitrary, according to \eqref{vd:2ch}, the variational derivative of $\lambda$ with respect to $m$ is given by
\begin{equation*}
	\delta_{m}\lambda = - \frac{\lambda f \hat{g}}{\int (m - 2\lambda\sigma \rho^{2})f\hat{g} \mathrm{d}x}.
\end{equation*}
Similarly, the variational derivative of $\lambda$ with respect to $\rho$ is obtained as
\begin{equation*}
	\delta_{\rho}\lambda = \frac{2\lambda^{2}\sigma \rho f\hat{g} }{\int (m - 2\lambda\sigma \rho^{2})f\hat{g} \mathrm{d}x}.
\end{equation*}
Hence, we have
\begin{lemma}\label{lem:2}
	Regarding the linear spectral problem \eqref{sp:2ch:a}\eqref{sp:2ch:b}, together with the adjoint one \eqref{adsp:2ch:a}\eqref{adsp:2ch:b}, the functional gradient of the spectral parameter $\lambda$ is 
	\begin{equation}\label{sg:2ch}
		\begin{pmatrix}
			\delta_{m}\lambda \\ \delta_{\rho}\lambda
		\end{pmatrix} = - \frac{1}{\int (m - 2\lambda\sigma \rho^{2})f\hat{g} \mathrm{d}x} \begin{pmatrix}
		    \lambda f \hat{g} \\
		    - 2\lambda^{2}\sigma \rho f\hat{g}
		\end{pmatrix} .
	\end{equation}
\end{lemma}
\begin{remark}
	The functional gradient of $\lambda$ will be used to generate nonlocal infinitesimal symmetries. The factor $ - \frac{1}{\int (m - 2\lambda\sigma \rho^{2})f\hat{g} \mathrm{d}x}$ in \eqref{sg:2ch} is not essential for this purpose, and will be discarded. 
\end{remark}

Applying the Hamiltonian operator $\mathcal{B}_1$ (please refer to \eqref{hamop:2ch}) to the functional gradient of $\lambda$ and simplifying by use of \eqref{sp:2ch:a} and \eqref{adsp:2ch:a}, we get 
\begin{align}
	\begin{pmatrix}
		\Omega^{m}\\\Omega^{\rho}	
	\end{pmatrix} \equiv &\; \left. \mathcal{B}_{1} \begin{pmatrix}
		\lambda f \hat{g} \\
		-2\lambda^{2}\sigma \rho f \hat{g}	
	\end{pmatrix} \right|_{\eqref{sp:2ch:a}\eqref{adsp:2ch:a}} \notag \\
	=&\; \begin{pmatrix}
		 4\lambda^2 \left(m - \lambda\sigma \rho^{2}\right)\left(g\hat{g} - f\hat{f}\right) - 4\lambda^{3}\sigma \rho\rho_{x}f\hat{g} + 2\lambda^{2} m_{x}f\hat{g} \\
		2\lambda^{2}\rho \left(g\hat{g} - f\hat{f}\right) + 2\lambda^{2}\rho_{x} f\hat g
	\end{pmatrix}. \label{mrhoinf:2ch}
\end{align} 
It is noted that the Hamiltonian operator $\mathcal{B}_2$ (please refer to \eqref{hamop:2ch}) maps the functional gradient of $\lambda$ to
\begin{equation*}
	\left. \mathcal{B}_{2} \begin{pmatrix}
		\lambda f \hat{g} \\
		-2\lambda^{2}\sigma  \rho f \hat{g}	
	\end{pmatrix} \right|_{\eqref{sp:2ch:a}\eqref{adsp:2ch:a}} = - \frac{1}{2\lambda}\begin{pmatrix}
	\Omega^{m}\\\Omega^{\rho}	
	\end{pmatrix}.
\end{equation*}

Suppose that the 2-CH equation \eqref{eq:2ch} is invariant (up to the first degree of $\varepsilon$) under an infinitesimal transformation
\begin{equation}\label{infsym:2ch}
	u\mapsto u + \varepsilon \Omega^u,\quad m\mapsto m + \varepsilon \Omega^m,\quad \rho\mapsto \rho + \varepsilon \Omega^{\rho},
\end{equation}
where $\Omega^m$ and $\Omega^{\rho}$ are defined by \eqref{mrhoinf:2ch}. Because of the relation $m = u - u_{xx}$ in the 2-CH equation \eqref{eq:2ch}, it should hold that $\Omega^{m} \equiv (\partial_x - \partial_x^3)\left(\lambda f\hat{g}\right) = \Omega^{u} - \Omega^{u}_{~xx}$. Hence, we take
\begin{equation} \label{uinf:2ch}
	\Omega^{u} \equiv \left.\partial_x\left(\lambda f\hat{g}\right)\right|_{\eqref{sp:2ch:a}\eqref{adsp:2ch:a}} = \lambda\left(g\hat{g} - f\hat{f}\right).
\end{equation}
By direct calculations, it is proven that
\begin{prop}\label{prop:1}
	The 2-CH equation \eqref{eq:2ch} is invariant (up to the first degree of $\varepsilon$) under an infinitesimal transformation \eqref{infsym:2ch}, where $\Omega^u$, $\Omega^{m}$ and $\Omega^{\rho}$ are defined by \eqref{uinf:2ch}\eqref{mrhoinf:2ch}. 
\end{prop}
In other words, \emph{$(\Omega^{u},\Omega^{m},\Omega^{\rho})$ defined by \eqref{uinf:2ch}\eqref{mrhoinf:2ch} is an infinitesimal symmetry of the 2-CH equation \eqref{eq:2ch}.}

According to lemma \ref{lem:1}, letting $(\hat{f},\hat{g}) = (-g,f)$ in \eqref{uinf:2ch}\eqref{mrhoinf:2ch}, we infer a reduced nonlocal infinitesimal symmetry  of the 2-CH equation \eqref{eq:2ch} from proposition \ref{prop:1}.
\begin{coro}\label{cor:2ch}
	The 2-CH equation \eqref{eq:2ch} is invariant (up to the first degree of $\varepsilon$) under the infinitesimal transformation
	\begin{equation}\label{redinfsym:2ch}
		u\mapsto u + \varepsilon \omega^u,\quad m\mapsto m + \varepsilon \omega^m,\quad \rho\mapsto \rho + \varepsilon \omega^{\rho},
	\end{equation}
	where $\omega^u$, $\omega^m$ and $\omega^\rho$ are given by
	\begin{align*}
		\begin{pmatrix}
			\omega^{u} \\ \omega^{m} \\ \omega^{\rho}
		\end{pmatrix} \equiv \frac{1}{2}\left.\begin{pmatrix}
		    \Omega^{u} \\ \Omega^{m} \\ \Omega^{\rho}
		\end{pmatrix}\right|_{(\hat{f},\hat{g}) = (-g,f)} = \begin{pmatrix}
		    \lambda f g \\
		    4\lambda^2 \left(m - \lambda\sigma \rho^{2}\right) fg- 2\lambda^3\sigma \rho\rho_{x}f^2 + \lambda^2 m_{x}f^2\\
		    2\lambda^2\rho fg + \lambda^2\rho_{x}f^2
		\end{pmatrix}.
	\end{align*}
\end{coro}

\subsection{Equation \eqref{eq:2mch}}
By the same method, nonlocal infinitesimal symmetries are constructed for equation \eqref{eq:2mch}. We skip some calculating details, and just state key conclusions. 

Equation \eqref{eq:2mch} is a bi-Hamiltonian system \cite{tianliu13}, i.e.
\begin{equation*}
	\begin{pmatrix}
		m_{t} \\ n_{t}
	\end{pmatrix}  = \mathcal{D}_{1} \begin{pmatrix}
		\delta_{m} \\ \delta_{n} 
	\end{pmatrix} \int m q r^{2} \mathrm{d}x = \mathcal{D}_{2} \begin{pmatrix}
		\delta_{m} \\ \delta_{n} 
	\end{pmatrix}  \int m r \mathrm{d}x,
\end{equation*}
where two compatible Hamiltonian operator $\mathcal{D}_1$ and $\mathcal{D}_2$ are given by 
\begin{equation}\label{hamop:2mch}
	\mathcal{D}_{1}=  \begin{pmatrix}
		0 & \partial_{x}^{2}-\partial_{x}\\
		-\partial_{x}^{2}-\partial_{x} & 0	
	\end{pmatrix} ,\quad \mathcal{D}_{2}= \begin{pmatrix}
		\partial_{x}m\partial^{-1}_{x}m\partial_{x}& 	\partial_{x}m\partial^{-1}_{x}n\partial_{x} \\
		\partial_{x}n\partial^{-1}_{x}m\partial_{x} & \partial_{x}n\partial^{-1}_{x}n\partial_{x}
	\end{pmatrix}.
\end{equation}
There is a linear spectral problem with the spectral parameter $z$ for equation \eqref{eq:2mch} \cite{sqq11,tianliu13}
\begin{subequations}  
	\begin{align}
		\begin{pmatrix}
			f \\g	
		\end{pmatrix}_{x} =&\; \mathbb{F}\begin{pmatrix}
		f \\g	
		\end{pmatrix}, \quad  \mathbb{F}\equiv \begin{pmatrix}
			\frac{1}{2} & zm \\
			zn& -\frac{1}{2}
		\end{pmatrix}, \label{sp:2mch:a}\\
		\begin{pmatrix}
			f \\g	
		\end{pmatrix}_{t} =&\; \mathbb{G} \begin{pmatrix}
		f \\g	
	    \end{pmatrix},\quad \mathbb{G} \equiv \begin{pmatrix}
			\frac{1}{4z^{2}}+\frac{1}{2}qr & \frac{q}{2z}+zmqr\\
			-\frac{r}{2z}+znqr &-\frac{1}{4z^{2}}-\frac{1}{2}qr
		\end{pmatrix}, \label{sp:2mch:b}
	\end{align}
\end{subequations}
and correspondingly the adjoint spectral problem is formulated as
\begin{subequations}  
	\begin{align}
		\begin{pmatrix}
			\hat{f} & \hat{g}	
		\end{pmatrix}_{x} =&\; - \begin{pmatrix}
			\hat{f} & \hat{g}	
		\end{pmatrix}\mathbb{F},\label{adsp:2mch:a}\\
		\begin{pmatrix}
			\hat{f} & \hat{g}	
		\end{pmatrix}_{t} =&\; - \begin{pmatrix}
			\hat{f} & \hat{g}	
		\end{pmatrix}\mathbb{G}. \label{adsp:2mch:b}
	\end{align}	
\end{subequations}

By calculating the variational derivatives of the spectral parameter $z$ with respect to $m$ ($n$ respectively), we obtain the functional gradient of $z$
\begin{equation}
	\begin{pmatrix}
		\delta_{m}z \\ \delta_{n}z	
	\end{pmatrix} = \frac{-z}{\int (mg\hat{f} + nf\hat{g} )\mathrm{d}x} \begin{pmatrix}
		 g\hat{f} \\ f\hat{g} 	
	\end{pmatrix}.
\end{equation}
Applying the Hamiltonian operator $\mathcal{D}_1$ (please refer to \eqref{hamop:2mch}) to the functional gradient of $z$ (without the factor $\frac{-z}{\int (mg\hat{f} + nf\hat{g} )\mathrm{d}x}$) and simplifying expressions with \eqref{sp:2mch:a} and \eqref{adsp:2mch:a} yield
\begin{equation} \label{mninf:2mch}
	\begin{pmatrix}
			\Omega^{m} \\ \Omega^{n}	
		\end{pmatrix} \equiv \left. \mathcal{D}_{1} \begin{pmatrix}
			g\hat{f}  \\  f\hat{g}	
		\end{pmatrix} \right|_{\eqref{sp:2mch:a}\eqref{adsp:2mch:a}} = \begin{pmatrix}
			2z^{2}m\left(nf\hat{g} - m g\hat{f}\right) - zm_{x}\left(f\hat{f}-g\hat{g}\right)\\
			2z^{2}n\left(n f\hat{g} - mg\hat{f}\right) - zn_{x}\left(f\hat{f}-g\hat{g}\right)
		\end{pmatrix}.
\end{equation}
Relations $m = q - q_{x}$ and $n = - r - r_{x}$ imply 
\begin{equation*}
	\Omega^{m} \equiv - (\partial_x - \partial_x^2)\left(f\hat{g}\right) = \Omega^{q} - \Omega^{q}_{~x},\quad \Omega^{n} \equiv (- \partial_x - \partial_x^2)\left(g\hat{f}\right) = - \Omega^{r} - \Omega^{r}_{~x}.
\end{equation*}
So it is reasonable to set 
\begin{align}
	\Omega^{q} \equiv&\; \left.-\partial_x \left(f\hat{g}\right)\right|_{\eqref{sp:2mch:a}\eqref{adsp:2mch:a}} = zm\left(f\hat{f} - g\hat{g}\right) - f\hat{g}, \label{qinf:2mch}\\
	\Omega^{r} \equiv&\; \left.\partial_x \left(g\hat{f}\right)\right|_{\eqref{sp:2mch:a}\eqref{adsp:2mch:a}} = zn\left(f\hat{f} - g\hat{g}\right) - g\hat{f}. \label{rinf:2mch}
\end{align}
\begin{prop}\label{prop:2}
	Equation \eqref{eq:2mch} is invariant (up to the first degree of $\varepsilon$) under the infinitesimal transformation
	\begin{equation}\label{infsym:2mch}
		q\mapsto q + \varepsilon \Omega^q,\quad r\mapsto r + \varepsilon \Omega^r,\quad m\mapsto m + \varepsilon \Omega^m,\quad n\mapsto n + \varepsilon \Omega^n,
	\end{equation}
	where $\Omega^q$, $\Omega^r$, $\Omega^m$ and $\Omega^n$ are defined by \eqref{qinf:2mch}\eqref{rinf:2mch} and \eqref{mninf:2mch}.
\end{prop}

Given any solution $(f,g)^{\top}$ to the linear spectral problem \eqref{sp:2mch:a}\eqref{sp:2mch:b}, then it is checked that $(-g,f)$ fulfils the adjoint spectral problem \eqref{adsp:2mch:a}\eqref{adsp:2mch:b}. So the infinitesimal symmetry \eqref{infsym:2mch} could be reduced.
\begin{coro}\label{cor:2mch}
	Equation \eqref{eq:2mch} is invariant (up to the first degree of $\varepsilon$) under the infinitesimal transformation
	\begin{equation}\label{redinfsym:2mch}
		q\mapsto q + \varepsilon \omega^q,\quad r\mapsto r + \varepsilon \omega^r,\quad m\mapsto m + \varepsilon \omega^m,\quad n\mapsto n + \varepsilon \omega^n,
	\end{equation}
	where $\omega^{q}$, $\omega^{r}$, $\omega^{m}$ and $\omega^{n}$ are given by
	\begin{equation*}
		\begin{pmatrix}
			\omega^{q} \\ \omega^{r} \\ \omega^{m} \\ \omega^{n}
		\end{pmatrix} \equiv \left.\begin{pmatrix}
			\Omega^{q} \\ \Omega^{r} \\ \Omega^{m} \\ \Omega^{n}
		\end{pmatrix}\right|_{(\hat{f},\hat{g}) = (-g,f)} = \begin{pmatrix}
			-2zmfg - f^2 \\
			-2znfg + g^2 \\
			2z^2m\left(nf^2 + mg^2\right) + 2zm_x fg\\
			2z^2n\left(nf^2 + mg^2\right) + 2zn_x fg
		\end{pmatrix}.
	\end{equation*}
\end{coro}

\begin{remark}\label{rem:2}
	As mentioned in Introduction, equation \eqref{eq:2mch} could be converted to equation \eqref{eq:sqq} by redefining $q = u_{x} + u$ and $r = v_{x} - v$. However at present, it is unclear how to prolong infinitesimal transformations \eqref{infsym:2mch} or \eqref{redinfsym:2mch} to $u$ and $v$. One may turn to other approaches for constructing nonlocal symmetries of equation \eqref{eq:sqq}.  
\end{remark}

\section{Finite symmetry transformations}\label{sec:3}
Nonlocal infinitesimal symmetries (please refer to corollaries \ref{cor:2ch} and \ref{cor:2mch} in section \ref{sec:2}) will be prolonged to enlarged systems associated with the 2-CH equation \eqref{eq:2ch} (equation \eqref{eq:2mch} respectively), and then integrated to yield symmetry transformations in finite form for enlarged systems. Moreover, nontrivial solutions will be derived for both equations. 

\subsection{For the 2-CH equation \eqref{eq:2ch}}\label{subsec:3:1}
Besides the 2-CH equation \eqref{eq:2ch} and its linear spectral problem \eqref{sp:2ch:a}\eqref{sp:2ch:b}, we introduce a first-order system
\begin{subequations}  
	\begin{align}
	p_{x} =&\; - \lambda^3\left(m - 2\lambda\sigma \rho^2\right)f^2, \label{p:2ch:a} \\
	p_{t} =&\; \lambda^3u\left(m - 2\lambda\sigma \rho^2\right)f^2 + \frac{\lambda}{8}\left(1 - 4\lambda^2\sigma\rho^2\right)f^2 - \frac{\lambda}{2}g^2, \label{p:2ch:b}
	\end{align}
\end{subequations}
which are checked to be compatible, and thus bring us with a new pseudo-potential $p$. We consider {\bf an enlarged system consisting of \eqref{eq:2ch}, \eqref{sp:2ch:a}\eqref{sp:2ch:b} and \eqref{p:2ch:a}\eqref{p:2ch:b}}, which involves dependent variables $u$, $m$, $\rho$, $f$, $g$ and $p$. Prolonging the nonlocal infinitesimal symmetry \eqref{redinfsym:2ch} in corollary \ref{cor:2ch} to the enlarged system yields an infinitesimal symmetry of the enlarged system, expressed in terms of an evolutionary vector field as
\begin{align*}
	\lambda fg \frac{\partial}{\partial u} &\; + \lambda\left(\left(\frac{1}{4} - \lambda m + \lambda^2\sigma\rho^2\right)f^2 + g^2 \right)\frac{\partial}{\partial u_x} + \lambda^2\left(2\rho fg + \rho_{x}f^2\right)\frac{\partial}{\partial \rho} \\
	&\; + \lambda^2\left(2\rho\left(\frac{1}{4} - \lambda m + \lambda^2\sigma\rho^2\right)f^2 + 2\left(\rho g + 2\rho_{x}f\right)g + \rho_{xx}f^2 \right)\frac{\partial}{\partial \rho_{x}} \\
	&\; + \lambda^2 \Big(4\left(m - \lambda\sigma\rho^{2}\right) fg - 2\lambda\sigma\rho\rho_{x}f^2 + m_{x}f^2\Big)\frac{\partial}{\partial m} + fp\frac{\partial}{\partial f} \\
	&\; + \Big(gp - \lambda^{3}\left(m - 2\lambda\sigma\rho^2\right)f^3\Big)\frac{\partial}{\partial g} + \Big(p^{2} + \lambda^{6}\sigma \rho^2f^{4} + \lambda^{2} p_{x} f^{2}\Big)\frac{\partial}{\partial p},
\end{align*}
which is equivalent to the non-evolutionary vector field
\begin{align}
	\bm{V} =&\; - \lambda^2 f^2\frac{\partial}{\partial x} + \Big(\lambda fg - \lambda^2 u_x f^2\Big) \frac{\partial}{\partial u} + \lambda\left(\left(\frac{1}{4} - \lambda u + \lambda^2\sigma\rho^2\right)f^2 + g^2 \right)\frac{\partial}{\partial u_x} \notag\\
	&\;  + 2\lambda^2\rho fg\frac{\partial}{\partial \rho} + \lambda^2\left(2\rho\left(\frac{1}{4} - \lambda m + \lambda^2\sigma\rho^2\right)f^2 + 2\left(\rho g + 2\rho_{x}f\right)g \right)\frac{\partial}{\partial \rho_{x}} \notag \\
	&\; + \lambda^2\Big(4 \left(m - \lambda\sigma\rho^{2}\right) fg- 2\lambda\sigma\rho\rho_{x}f^2 \Big)\frac{\partial}{\partial m} + \Big(fp - \lambda^2f^2g\Big) \frac{\partial}{\partial f} \notag \\
	&\; + \left(gp - \lambda^{2}\left(\frac{1}{4} - \lambda^2\sigma\rho^{2}\right)f^{3}\right)\frac{\partial}{\partial g} + \Big(p^{2} + \lambda^{6}\sigma\rho^{2} f^{4}\Big)\frac{\partial}{\partial p}.\label{pvf:2ch}
\end{align}
It should be remarked that $\bm{V}$ is understood as a vector field in the space with coordinates $(x,t,u,u_{x},m,\rho,\rho_{x},f,g,p)$, where $u_{x}$ and $\rho_{x}$ play the same roles as other dependent variables.  

The vector field $\bm{V}$ defined by \eqref{pvf:2ch} generates the one-parameter symmetry group of the enlarged system \eqref{eq:2ch}\eqref{sp:2ch:a}\eqref{sp:2ch:b}\eqref{p:2ch:a}\eqref{p:2ch:b}
\begin{equation}\label{nvar:2ch}
	\left(\widetilde{x},\widetilde{t},\widetilde{u},\widetilde{m},\widetilde{\rho},\widetilde{f},\widetilde{g},\widetilde{p}\right)\equiv \exp(\varepsilon\bm{V})\left(x,t,u,m,\rho,f,g,p\right),
\end{equation}
which is explicitly formulated as
\begin{subequations}
	\begin{align}
		\widetilde{x} =&\; x + \ln \left|\frac{2(1-\varepsilon p+\varepsilon\lambda^2fg) - \varepsilon \lambda^2 f^2}{2(1-\varepsilon p+\varepsilon\lambda^2fg) + \varepsilon \lambda^2 f^2}\right|,\qquad \widetilde{t} = t, \label{nx:2ch}\\
		\widetilde{u} =&\; \frac{u\left[4(1-\varepsilon p+\varepsilon\lambda^2fg)^2 + \varepsilon^2\lambda^4 f^4\right] - 4\varepsilon\lambda(\lambda u_{x}f^2 - fg)(1-\varepsilon p+\varepsilon\lambda^2fg) - \varepsilon^2\lambda^3f^4 }{4(1-\varepsilon p+\varepsilon\lambda^2fg)^2 - \varepsilon^2\lambda^4 f^4} , \label{nu:2ch}\\
		\widetilde{m} =&\; \frac{\left[4(1-\varepsilon p+\varepsilon\lambda^2fg)^2 - \varepsilon^2\lambda^4 f^4\right]^2}{16\left[(1-\varepsilon p)^2 - \varepsilon^2\lambda^6\sigma \rho^2f^4\right]^2}\left[m \frac{(1-\varepsilon p)^2 + \varepsilon^2\lambda^6\sigma \rho^2f^4}{(1-\varepsilon p)^2 - \varepsilon^2\lambda^6\sigma \rho^2f^4} \right.\notag \\
		&\;\qquad\qquad\qquad\qquad\qquad\qquad \left. - \frac{2\varepsilon\lambda^3\sigma\left(\rho\rho_{x}f^2 + 2\rho^2fg\right)(1-\varepsilon p) + 4\varepsilon^2\lambda^7\sigma^2\rho^4f^4}{(1-\varepsilon p)^2 - \varepsilon^2\lambda^6\sigma \rho^2f^4} \right], \label{nm:2ch}\\
		\widetilde{\rho} =&\; \frac{\rho\left[4(1-\varepsilon p+\varepsilon\lambda^2fg)^2 - \varepsilon^2\lambda^4 f^4\right]}{4\left[(1-\varepsilon p)^2 - \varepsilon^2\lambda^6\sigma \rho^2 f^4\right]}, \label{nrho:2ch}\\
		\widetilde{f} =&\; \frac{2f}{\sqrt{4(1-\varepsilon p+\varepsilon\lambda^2fg)^2 - \varepsilon^2\lambda^4 f^4}}, \notag \\
		\widetilde{g} =&\; \frac{4g\left[(1-\varepsilon p)^2 + \varepsilon^2\lambda^{6}\sigma\rho^{2} f^4\right] - \varepsilon\lambda^2f(1-\varepsilon p)\left(f^2-4g^2-4\lambda^2\sigma \rho^2 f^2\right)}{2\left[(1-\varepsilon p)^2 - \varepsilon^2\lambda^6\sigma\rho^{2} f^{4}\right]\sqrt{4(1-\varepsilon p+\varepsilon\lambda^2 fg)^2 - \varepsilon^2\lambda^4 f^4}}, \notag \\
		\widetilde{p} =&\; \frac{p(1-\varepsilon p) + \varepsilon \lambda^6\sigma \rho^{2}f^{4}}{(1-\varepsilon p)^2 - \varepsilon^2\lambda^6 \sigma\rho^{2} f^{4}} . \notag
	\end{align}
\end{subequations}
Through direct but tedious calculations, it is shown that \emph{the enlarged system \eqref{eq:2ch}\eqref{sp:2ch:a}\eqref{sp:2ch:b}\eqref{p:2ch:a}\eqref{p:2ch:b} is invariant under the transformation \eqref{nvar:2ch}.} Specially, when the enlarged system holds, there exists a 2-CH equation in terms of new variables $(\widetilde{x},\widetilde{t},\widetilde{u},\widetilde{m})$ defined by \eqref{nx:2ch}-\eqref{nrho:2ch}, i.e.
\begin{equation*}
	\left\{	\begin{aligned}
		\widetilde{m}_{\widetilde{t}} =&\; -\widetilde{u}\widetilde{m}_{\widetilde{x}}-2\widetilde{u}_{\widetilde{x}}\widetilde{m}+\sigma^{2}\widetilde{\rho}\widetilde{\rho}_{\widetilde{x}} , \quad \widetilde{m} = \widetilde{u} - \widetilde{u}_{\widetilde{x}\widetilde{x}}, \\
		\widetilde{\rho}_{\widetilde{t}} =&\; -(\widetilde{\rho} \widetilde{u})_{\widetilde{x}} .
	\end{aligned} \right. 
\end{equation*}

With the finite symmetry transformations \eqref{nvar:2ch}, we shall construct nontrivial solutions to the 2-CH equation \eqref{eq:2ch}. For instance, given a trivial solution $(u,m,\rho)=(u_{0},u_{0},\rho_{0})$ of the 2-CH equation \eqref{eq:2ch}, where $u_{0}$ and $\rho_{0}$ are assumed to be constants such that 
\begin{equation*}
	u_{0}^2 > \sigma\rho_{0}^2\quad \text{and}\quad \frac{1}{4}-\lambda u_{0}+\lambda^{2}\sigma\rho_{0}^{2}>0.
\end{equation*}
then we obtain a special solution to \eqref{sp:2ch:a}\eqref{sp:2ch:b} (with $(u,m,\rho)=(u_{0},u_{0},\rho_{0})$)
\begin{align*}
	\begin{pmatrix}
		f \\ g
	\end{pmatrix} = \begin{pmatrix}
	    1 \\ k
	\end{pmatrix} \exp \left(k\left[x-\left(\frac{1}{2\lambda}+u_{0}\right)t\right]\right),
\end{align*}
and correspondingly a solution to \eqref{p:2ch:a}\eqref{p:2ch:b}
\begin{align*}
	p = \frac{\lambda^3}{2k}\left(\lambda\sigma\rho_{0}^2 - u_{0}\right)\exp \left(2k\left[x-\left(\frac{1}{2\lambda}+u_{0}\right)t\right]\right),
\end{align*}
where $k=\sqrt{\frac{1}{4}-\lambda u_{0}+\lambda^{2}\sigma\rho_{0}^{2}}$ (just the positive square root). Substituting them into \eqref{nx:2ch}-\eqref{nrho:2ch} yields a nontrivial solution to the 2-CH equation \eqref{eq:2ch} in parametric expressions
\begin{align*}
	\widetilde{x} =&\; x+\ln \left|1-2\lambda u_{0}-2k\Xi \right|-\ln\left|1-2\lambda u_{0}+2k \right |, \qquad \widetilde{t} = t,\\
	\widetilde{u} =&\; \frac{1+2\lambda u_{0}+2k\Xi}{4\lambda}-\frac{\lambda(u_{0}^{2}-\sigma\rho_{0}^{2})}{1-2\lambda u_{0}-2k\Xi},\\
	\widetilde{m} =&\; \frac{1}{\left(\frac{1+2\lambda u_{0}-2k\Xi}{2}+\frac{2\lambda^{2}(u_{0}^{2}-\sigma\rho_{0}^{2})}{1-2\lambda u_{0}-2k\Xi}\right)^2}\left[u_{0} - 3\lambda\sigma\rho_{0}^{2} + 8\lambda\sigma\rho_{0}^{2}\left(\frac{2\lambda^{2}(u_{0}^{2}-\sigma\rho_{0}^{2})}{2k\Xi+2\lambda u_{0}-1} + k\Xi - \lambda u_{0}\right)\right], \\
	\widetilde{\rho} =&\; \frac{\rho_{0}}{\frac{1+2\lambda u_{0}-2k\Xi}{2}+\frac{2\lambda^{2}(u_{0}^{2}-\sigma\rho_{0}^{2})}{1-2\lambda u_{0}-2k\Xi}},
\end{align*}
where 
\begin{equation*}
	\Xi \equiv \left\{\begin{aligned}
		& \tanh\left(k\left[x-\left(\frac{1}{2\lambda}+u_{0}\right)t\right] +  \ln\sqrt{\frac{\varepsilon\lambda^{2}}{4k}\left(1 - 2\lambda u_{0} + 2k\right)}\right),\quad  & \left(\varepsilon > 0 \right), \\
		& \coth\left(k\left[x-\left(\frac{1}{2\lambda}+u_{0}\right)t\right] +  \ln\sqrt{\frac{(-\varepsilon)\lambda^{2}}{4k}\left(1 - 2\lambda u_{0} + 2k\right)}\right),\quad  & \left(\varepsilon < 0\right).
	\end{aligned}\right.
\end{equation*}

\subsection{For equation \eqref{eq:2mch}}
In addition to equation \eqref{eq:2mch} and its linear spectral problem \eqref{sp:2mch:a}\eqref{sp:2mch:b}, a first order system with a new pseudo-potential $p$ is introduced as
\begin{subequations}  
	\begin{align}
		p_{x} =&\; -z^{2}nf^{2}, \label{px:2mch}\\
		p_{t} =&\; -z^{2}nqrf^{2}-\frac{1}{2}qg^{2}-\frac{fg}{2z}. \label{pt:2mch}
	\end{align}
\end{subequations}
The nonlocal infinitesimal symmetry \eqref{redinfsym:2mch} in corollary \ref{cor:2mch} is prolonged to {\bf the enlarged system \eqref{eq:2mch}\eqref{sp:2mch:a}\eqref{sp:2mch:b}\eqref{px:2mch}\eqref{pt:2mch}}, and consequently we get an infinitesimal symmetry of the enlarged system
\begin{align*}
	- \left(2zmfg + f^2\right) \frac{\partial}{\partial q} &\; - \left(2znfg - g^2\right)\frac{\partial}{\partial r} + \left(2z^2m(nf^2 + mg^2) + 2zm_{x}fg\right)\frac{\partial}{\partial m} \\
	&\;  + \left(2z^2n(nf^2 + mg^2) + 2zn_{x}fg\right)\frac{\partial}{\partial n} + \left(fp + 2z f_{x}fg\right)\frac{\partial}{\partial f} \\
	&\; + \left(gp + zfg^2 + 2zg_{x}fg\right)\frac{\partial}{\partial g} + \left(p^2 + 2zp_{x}fg\right)\frac{\partial}{\partial p},
\end{align*}
or equivalently in the non-evolutionary form
\begin{align*}
	\bm{W} =&\; - 2zfg\frac{\partial}{\partial x} - \left(2zqfg + f^2\right) \frac{\partial}{\partial q} + \left(2zrfg + g^2\right)\frac{\partial}{\partial r} + 2z^2m\left(nf^2 + mg^2\right)\frac{\partial}{\partial m}  \\
	&\;  + 2z^2n\left(nf^2 + mg^2\right)\frac{\partial}{\partial n} + fp\frac{\partial}{\partial f} + \left(gp + zfg^2\right)\frac{\partial}{\partial g} + p^2\frac{\partial}{\partial p}.
\end{align*}
The one-parameter transformation generated by the vector field $\bm{W}$, i.e.
\begin{equation}\label{nvar:2mch}
	\left(\widetilde{x},\widetilde{t},\widetilde{q},\widetilde{r},\widetilde{m},\widetilde{n},\widetilde{f},\widetilde{g},\widetilde{p}\right)\equiv \exp(\varepsilon\bm{W})\left(x,t,q,r,m,n,f,g,p\right),
\end{equation}
is explicitly formulated as
\begin{subequations}
	\begin{align}
		\widetilde{x} =&\; x + \ln \frac{(1 - \varepsilon p - \varepsilon zfg)^2}{(1 - \varepsilon p)^2}, \qquad \widetilde{t} = t, \label{nxt:2mch}\\
		\widetilde{q} =&\; q\frac{(1 - \varepsilon p - \varepsilon zfg)^2}{(1 - \varepsilon p)^2} - \frac{\varepsilon f^2 (1 - \varepsilon p - \varepsilon zfg)}{(1 - \varepsilon p)^2} ,\label{nq:2mch} \\
		\widetilde{r} =&\; r\frac{(1 - \varepsilon p)^2}{(1 - \varepsilon p - \varepsilon zfg)^2} + \frac{\varepsilon g^2(1 - \varepsilon p)}{(1 - \varepsilon p - \varepsilon zfg)^2}, \label{nr:2mch} \\
		\frac{1}{\widetilde{m}} = &\; \frac{1}{m} - \frac{2\varepsilon z^2 n f^2}{m(1 - \varepsilon p)} - \frac{2\varepsilon z^2 g^2}{1 - \varepsilon p - \varepsilon zfg} , \label{nm:2mch}\\
		\frac{1}{\widetilde{n}} = &\; \frac{1}{n} - \frac{2\varepsilon z^2 m g^2}{n(1 - \varepsilon p - \varepsilon zfg)} - \frac{2\varepsilon z^2 f^2}{1 - \varepsilon p}, \label{nn:2mch} \\
		\widetilde{f} =&\; \frac{f}{1 - \varepsilon p},\qquad \widetilde{g} = \frac{g}{1 - \varepsilon p - \varepsilon zfg},\qquad \widetilde{p} = \frac{p}{1 - \varepsilon p}. \notag
	\end{align}
\end{subequations}
By direct calculations, \emph{the enlarged system \eqref{eq:2mch}\eqref{sp:2mch:a}\eqref{sp:2mch:b}\eqref{px:2mch}\eqref{pt:2mch} is shown to be invariant under the transformation \eqref{nvar:2mch}.} It should be emphasized that new variables $(\widetilde{x},\widetilde{t},\widetilde{q},\widetilde{r},\widetilde{m},\widetilde{n})$ defined by \eqref{nxt:2mch}-\eqref{nn:2mch} satisfy
\begin{equation*}
	\left\{	\begin{aligned}
		\widetilde{m}_{\widetilde{t}} =&\; (\widetilde{m}\widetilde{q}\widetilde{r})_{\widetilde{x}}, &&  \widetilde{m} = \widetilde{q} - \widetilde{q}_{\widetilde{x}} ,\\
		\widetilde{n}_{\widetilde{t}} =&\; (\widetilde{n}\widetilde{q}\widetilde{r})_{\widetilde{x}}, &&  \widetilde{n} = - \widetilde{r} - \widetilde{r}_{\widetilde{x}} ,
	\end{aligned} \right.
\end{equation*}
when the  enlarged system \eqref{eq:2mch}\eqref{sp:2mch:a}\eqref{sp:2mch:b}\eqref{px:2mch}\eqref{pt:2mch} holds.

Given a trivial solution $(q,r,m,n) = (1,r_0,1,-r_{0})$ of \eqref{eq:2mch}, where $r_{0}$ is a constant such that 
\begin{align*}
	r_{0}>0,\quad \text{and}\quad \frac{1}{4}-z^{2}r_{0}>0.
\end{align*}
A solution of \eqref{sp:2mch:a}\eqref{sp:2mch:b} with $(q,r,m,n) = (1,r_0,1,-r_{0})$ is taken as
\begin{align*}
	\begin{pmatrix}
		f \\ g
	\end{pmatrix} = \begin{pmatrix}
	    1  \\ \frac{2k-1}{2z}
	\end{pmatrix}\exp{\left(k\left[x+\frac{(3-4k^{2})t}{4z^{2}}\right]\right)} ,
\end{align*}
while correspondingly a solution of \eqref{px:2mch}\eqref{pt:2mch} is 
\begin{align*}
	p=\frac{z^{2}r_{0}}{2k}\exp{\left(2k\left[x+\frac{(3-4k^{2})t}{4z^{2}}\right]\right)},
\end{align*}
where $k=\sqrt{\frac{1}{4}-z^{2}r_{0}}$.
Substituting them into \eqref{nxt:2mch}-\eqref{nn:2mch} yields a nontrivial solution of \eqref{eq:2mch}, i.e.
\begin{align*}
	& \widetilde{x} = x+2\ln\left |1-2k\Theta\right |-2\ln(1+2k), && \widetilde{t} = t,\\
	& \widetilde{q} = \left(1-\frac{1}{4z^{2}r_{0}}\right)\Theta^{2}+\frac{1}{4z^{2}r_{0}}, && \widetilde{m} = \frac{-1}{2k\Theta - \frac{4z^{2}r_{0}}{1-2k\Theta}},\\
	& \widetilde{r} = -\frac{4z^{2}r_{0}(1+2k-z^{2}r_{0})}{(1-2k\Theta)^{2}}+\frac{2z^{2}r_{0}}{1-2k\Theta}, 
	&& \widetilde{n} = \frac{r_{0}}{2k\Theta-\frac{4z^{2}r_{0}^{2}}{1-2k\Theta}},
\end{align*}
where 
\begin{align*}
	\Theta \equiv \left\{\begin{aligned}
		& \coth \left(k\left[x+\frac{(3-4k^{2})t}{4z^{2}}\right]+\ln \sqrt{\frac{\varepsilon(1-4k^{2})}{8k}}\right),\quad & \left(\varepsilon >0\right), \\
		& \tanh \left(k\left[x+\frac{(3-4k^{2})t}{4z^{2}}\right]+\ln\sqrt{\frac{(-\varepsilon)(1-4k^{2})}{8k}}\right),\quad & \left(\varepsilon <0\right).
	\end{aligned}\right.
\end{align*}

\begin{remark}\label{rem:3}
	As mentioned in Introduction, let
	\begin{equation}\label{cons:2mch}
		q = u_{x} + u,\quad r = - u_{x} + u,
	\end{equation}
	then $m = - n = u - u_{xx}$, and equation \eqref{eq:2mch} is reduced to the modified CH equation \eqref{eq:mch}. In accordance with the constraint \eqref{cons:2mch}, reducing nonlocal symmetries of equation \eqref{eq:2mch} would yield nonlocal symmetries of the modified CH equation \eqref{eq:mch}. Some results are briefly summarized as follows. An enlarged system consisting of the modified CH equation \eqref{eq:mch}, its linear spectral problem 
		\begin{subequations}  
			\begin{align}
				\begin{pmatrix}
					f \\g	
				\end{pmatrix}_{x} =&\; \begin{pmatrix}
					\frac{1}{2} & zm \\
					- zm & -\frac{1}{2}
				\end{pmatrix}\begin{pmatrix}
					f \\g	
				\end{pmatrix}, \label{sp:mch:a}\\
				\begin{pmatrix}
					f \\g	
				\end{pmatrix}_{t} =&\; \begin{pmatrix}
					\frac{1}{4z^{2}} - \frac{1}{2}(u_{x}^2 - u^2) & \frac{1}{2z}(u_{x} + u) - zm(u_{x}^2 - u^2) \\
					\frac{1}{2z}(u_{x} - u) - zm(u_{x}^2 - u^2) & -\frac{1}{4z^{2}} + \frac{1}{2}(u_{x}^2 - u^2)
				\end{pmatrix} \begin{pmatrix}
					f \\g	
				\end{pmatrix},\label{sp:mch:b}
			\end{align}
		\end{subequations}
		and a pair of compatible equations
		\begin{subequations}  
			\begin{align}
				p_{x} =&\; z^{2}mf^{2}, \label{p:mch:a}\\
				p_{t} =&\; - z^{2}m(u_{x}^2 - u^2)f^{2}-\frac{1}{2}(u_{x} + u)g^{2}-\frac{fg}{2z}, \label{p:mch:b}
			\end{align}
		\end{subequations} 
		admits an infinitesimal symmetry, expressed in terms of a non-evolutionary vector field as
		\begin{align*}
			\bm{W}_{\text{mCH}} =&\; - 2zfg\frac{\partial}{\partial x} - \frac{1}{2}\left(f^2 - g^2 + 4zu_{x}fg\right)\frac{\partial}{\partial u} - \frac{1}{2}\left(f^2 + g^2 + 4zufg\right)\frac{\partial}{\partial u_{x}} \\
			&\; - 2z^2m^2\left(f^2 - g^2\right)\frac{\partial}{\partial m} + fp\frac{\partial}{\partial f} + \left(gp + zfg^2\right)\frac{\partial}{\partial g} + p^2\frac{\partial}{\partial p},
		\end{align*}
		which generates the one-parameter symmetry transformations of the enlarged system (consisting of the modified CH equation \eqref{eq:mch}, \eqref{sp:mch:a}\eqref{sp:mch:b} and \eqref{p:mch:a}\eqref{p:mch:b}),
		\begin{equation*}
			\left(\widetilde{x},\widetilde{t},\widetilde{u},\widetilde{m},\widetilde{f},\widetilde{g},\widetilde{p}\right)\equiv \exp(\varepsilon\bm{W}_{\text{mCH}})\left(x,t,u,m,f,g,p\right),
		\end{equation*}
		explicitly given by 
		\begin{subequations}
			\begin{align}
				\widetilde{x} =&\; x + \ln \frac{(1 - \varepsilon p - \varepsilon zfg)^2}{(1 - \varepsilon p)^2}, \qquad \widetilde{t} = t, \label{nxt:mch}\\
				\widetilde{u} =&\; \left(u_{x} + u\right)\frac{(1 - \varepsilon p - \varepsilon zfg)^2}{2(1 - \varepsilon p)^2} - \left(u_{x} - u\right)\frac{(1 - \varepsilon p)^2}{2(1 - \varepsilon p - \varepsilon zfg)^2} \notag \\
				&\;\qquad\qquad\qquad\qquad - \frac{\varepsilon f^2 (1 - \varepsilon p - \varepsilon zfg)}{2(1 - \varepsilon p)^2} + \frac{\varepsilon g^2(1 - \varepsilon p)}{2(1 - \varepsilon p - \varepsilon zfg)^2} , \label{nu:mch}\\
				\frac{1}{\widetilde{m}} =&\; \frac{1}{m} + \frac{2\varepsilon z^2 f^2}{1 - \varepsilon p} - \frac{2\varepsilon z^2 g^2}{1 - \varepsilon p - \varepsilon zfg} , \label{nm:mch} \\
				\widetilde{f} =&\; \frac{f}{1 - \varepsilon p},\qquad \widetilde{g} = \frac{g}{1 - \varepsilon p - \varepsilon zfg},\qquad \widetilde{p} = \frac{p}{1 - \varepsilon p}.  \notag
			\end{align}
		\end{subequations}
\end{remark}

\section{B\"{a}cklund transformations} \label{sec:4}
Finite symmetry transformations presented in section \ref{sec:3} will lead to B\"{a}cklund transformations for the 2-CH equation \eqref{eq:2ch}, as well as equation \eqref{eq:2mch}. 

\subsection{A B\"{a}cklund transformation of the 2-CH equation \eqref{eq:2ch}} \label{sec:4:1}
About the finite symmetry transformation \eqref{nvar:2ch}, let us introduce 
\begin{equation}\label{s:2ch}
	s \equiv \frac{\varepsilon\lambda f^2}{1 - \varepsilon p},
\end{equation}
then in accordance with \eqref{sp:2ch:a} and \eqref{p:2ch:a}, we get
\begin{equation*}
	s_{x} + \lambda^2(m - 2\lambda \sigma \rho^2)s^2 = \frac{2\varepsilon\lambda fg}{1 - \varepsilon p}.
\end{equation*}
Moreover, from \eqref{sp:2ch:a}\eqref{sp:2ch:b} and \eqref{p:2ch:a}\eqref{p:2ch:b} we deduce that $s$ fulfils
\begin{subequations}  
	\begin{align}
		s_{xx} =&\; \frac{s_{x}^{2}}{2s} - \lambda^2\left[\left(m - 2\lambda\sigma\rho^2\right)s^2\right]_{x} - \frac{\lambda^4}{2}\left(m - 2\lambda\sigma\rho^2\right)^2s^3 + 2\left(\frac{1}{4}-\lambda m + \lambda^2\sigma\rho^2\right)s,  \label{sxx:2ch}\\
		s_{t} =&\; -\frac{1}{8}\left[s_{x} + \lambda^2(m - 2\lambda \sigma \rho^2)s^2\right]^2 - \left[s_{x} + \lambda^2(m - 2\lambda \sigma \rho^2)s^2\right]\left(u + \frac{1}{2\lambda}\right) \notag \\
		&\; + \left[\lambda^2u\left(m - 2\lambda\sigma\rho^2\right) - \frac{\lambda^2}{2}\sigma\rho^2 + \frac{1}{8}\right]s^2 + u_{x}s . \label{st:2ch}
	\end{align}
\end{subequations}
It is checked that \emph{\eqref{sxx:2ch} and \eqref{st:2ch} are compatible if and only if the 2-CH equation \eqref{eq:2ch} holds.} In terms of $s$ defined by \eqref{s:2ch}, \eqref{nx:2ch}-\eqref{nrho:2ch} could be reformulated as 
\begin{subequations}
	\begin{align}
		\widetilde{x} =&\; x + \ln \left| \frac{2 + \lambda s_{x} + \lambda^3(m - 2\lambda \sigma \rho^2)s^2 - \lambda s}{2 + \lambda s_{x} + \lambda^3(m - 2\lambda \sigma \rho^2)s^2 + \lambda s}\right|, \qquad \widetilde{t} = t, \label{btxt:2ch}\\
		\widetilde{u} =&\;  \frac{u \left(\left[2 + \lambda s_{x} + \lambda^3(m - 2\lambda \sigma \rho^2)s^2\right]^2 + \lambda^2 s^2\right) - \lambda s^2}{\left[2 + \lambda s_{x} + \lambda^3(m - 2\lambda \sigma \rho^2)s^2\right]^2 - \lambda^2 s^2} \notag \\
		&\; - \frac{\left[2\lambda u_{x} s - s_{x} - \lambda^2(m - 2\lambda \sigma \rho^2)s^2\right]\left[2 + \lambda s_{x} + \lambda^3(m - 2\lambda \sigma \rho^2)s^2\right]}{\left[2 + \lambda s_{x} + \lambda^3(m - 2\lambda \sigma \rho^2)s^2\right]^2 - \lambda^2 s^2} , \label{btu:2ch}\\
		\widetilde{m} =&\; \frac{\left(\left[2 + \lambda s_{x} + \lambda^3(m - 2\lambda \sigma \rho^2)s^2\right]^2 - \lambda^2 s^2\right)^2}{16\left[1 - \lambda^4\sigma\rho^2 s^2\right]^2}\left[m - \frac{2\lambda^2\sigma\left(\rho\rho_{x}s + \rho^2 s_{x}\right)}{1 - \lambda^4\sigma\rho^2 s^2} \right], \label{btm:2ch}\\
		\widetilde{\rho} =&\; \rho \frac{\left[2 + \lambda s_{x} + \lambda^3(m - 2\lambda \sigma \rho^2)s^2\right]^2 - \lambda^2 s^2}{4\left[1 - \lambda^4\sigma\rho^2 s^2\right]}. \label{btrho:2ch}
	\end{align}
\end{subequations}
Therefore, we have a B\"{a}cklund transformation for the 2-CH equation \eqref{eq:2ch}, which is summarized as follows.
\begin{prop}\label{prop:3}
	Assume that the 2-CH equation \eqref{eq:2ch}, as well as \eqref{sxx:2ch}\eqref{st:2ch}, holds, then $(\widetilde{x},\widetilde{t},\widetilde{u},\widetilde{m},\widetilde{\rho})$ defined by \eqref{btxt:2ch}-\eqref{btrho:2ch} gives us a solution in parametric expressions to the 2-CH equation \eqref{eq:2ch}.  
\end{prop}

\begin{remark}
	If we take
	\begin{equation*}
		\eta \equiv \frac{2 + \lambda s_{x} + \lambda^3(m - 2\lambda \sigma \rho^2)s^2 - \lambda s}{2 + \lambda s_{x} + \lambda^3(m - 2\lambda \sigma \rho^2)s^2 + \lambda s},
	\end{equation*}
	where $s$ is defined by \eqref{s:2ch}, then according to \eqref{sxx:2ch}\eqref{st:2ch}, it is proved that $\eta$ satisfies
	\begin{subequations}\label{etaxt:2ch}
		\begin{align}
			\eta_{x} = &\; 16\left(u_{x}^{2}+2uu_{x}+u^{2}-\sigma\rho^{2}\right)\eta^{2} + 2\lambda (u_{x} + u)\eta - \eta + \frac{\lambda^{2}}{16}, \label{etax:2ch}\\
			\eta_{t} = &\; -8\left[(u_{x}^{2}+u^{2}-\sigma\rho^{2})\left(2u+\frac{1}{\lambda}\right)-\frac{1}{\lambda}\left(2uu_{x} + u^{2} - \frac{4}{3}\lambda u^{3}\right)_{x} - \frac{2}{\lambda}(u+u_{x})_{t}\right]\eta^{2} \notag \\
			&\; - 2\lambda (u_{x} +  u)u\eta + \frac{1}{2\lambda}\eta - \frac{\lambda^2}{16}u - \frac{\lambda}{32}. \label{etat:2ch}
		\end{align}
	\end{subequations}
	In addition, equations \eqref{btxt:2ch}, \eqref{btu:2ch} and \eqref{btrho:2ch} are rewritten in terms of $\eta$ as
	\begin{subequations}\label{nbt:2ch}
		\begin{align}
			\widetilde{x} =&\; x + \ln \left| \eta \right|, \qquad \widetilde{t} = t, \label{nbtxt:2ch}\\
			\widetilde{u} =&\; u\left(1 + \frac{\eta_{x}}{\eta}\right) + \frac{\eta_{t}}{\eta},   \label{nbtu:2ch} \\
			\widetilde{\rho} =&\; \frac{\rho}{1 + \eta_{x}/\eta}. \label{nbtrho:2ch}
		\end{align}
	\end{subequations}
	When $\sigma = 1$, expressions \eqref{nbtxt:2ch}-\eqref{nbtrho:2ch} with $\eta$ determined by \eqref{etax:2ch}\eqref{etat:2ch} are nothing but the B\"{a}cklund transformation constructed in \cite{wangThesis21} through a reciprocal transformation converting  the 2-CH equation \eqref{eq:2ch} ($\sigma = 1$) to the associated 2-component CH equation.
\end{remark}

\begin{remark}
	When $\rho = 0$, results presented in subsections \ref{subsec:2:1} and \ref{subsec:3:1}, including the infinitesimal symmetry in corollary \ref{cor:2ch}, the finite symmetry transformation in \eqref{nvar:2ch} and the B\"{a}cklund transformation in proposition \ref{prop:3}, could be reduced to the case of the CH equation \eqref{eq:CH}. Especially, the transformations $x\mapsto \widetilde{x}$ and $m\mapsto \widetilde{m}$ given by \eqref{nx:2ch} and \eqref{nm:2ch} at $\rho = 0$ essentially coincide with results about the CH equation \eqref{eq:CH} in \cite{litian22}.
\end{remark}

\subsection{A B\"{a}cklund transformation of equation \eqref{eq:2mch}} \label{sec:4:2}
Inferred from the finite symmetry transformation \eqref{nvar:2mch}, a B\"{a}cklund transformation would be derived for equation \eqref{eq:2mch}. Let 
\begin{equation*}
	s \equiv \frac{\varepsilon z f^2}{1 - \varepsilon p},
\end{equation*}
then according to \eqref{sp:2mch:a}\eqref{sp:2mch:b} and \eqref{px:2mch}\eqref{pt:2mch}, we have 
\begin{equation*}
	\frac{\varepsilon z fg}{1 - \varepsilon p} = \frac{s_{x} + zn s^2 - s}{2zm},\quad 
	\frac{\varepsilon z g^2}{1 - \varepsilon p} = \frac{\left(s_{x} + zn s^2 - s\right)^2}{4z^2m^2s},
\end{equation*}
and 
\begin{subequations}  
	\begin{align}
		s_{xx} =&\; \frac{s_{x}^2}{2s} - z\left(ns^2\right)_{x} + \frac{m_{x}}{m}\left(s_{x} + zns^2 -s \right) - \frac{z^2}{2}n^2 s^3 + \left(2z^2mn + \frac{1}{2}\right)s, \label{sxx:2mch}\\
		s_{t} =&\; - \frac{q}{8z^3m^2}\left(s_{x} + zns^2 -s \right)^2 - \frac{\left(s - 2zq\right)}{4z^3m}\left(s_{x} + zns^2 -s \right) + qr s_{x} + \frac{s}{2z^2}. \label{st:2mch}
	\end{align}
\end{subequations}
It is checked that \emph{\eqref{sxx:2mch} and \eqref{st:2mch} are compatible if and only if equation \eqref{eq:2mch} holds}. Subsequently, \eqref{nxt:2mch}-\eqref{nn:2mch} are rewritten in terms of $s$ as
\begin{subequations}
	\begin{align}
		\widetilde{x} =&\; x + \ln \frac{\left[2zm - (s_{x} + zns^2 - s)\right]^2}{4z^2m^2},\qquad \widetilde{t} = t, \label{btxt:2mch}\\
		\widetilde{q} =&\; q \frac{\left[2zm - (s_{x} + zns^2 - s)\right]^2}{4z^2m^2} - \frac{\left[2zm - (s_{x} + zns^2 - s)\right]s}{2z^2m}, \label{btp:2mch}\\
		\widetilde{r} =&\; r \frac{4z^2m^2}{\left[2zm - (s_{x} + zns^2 - s)\right]^2} + \frac{(s_{x} + zns^2 - s)^2}{zs\left[2zm - (s_{x} + zns^2 - s)\right]^2}, \label{btr:2mch}\\
		\frac{1}{\widetilde{m}} =&\; \frac{1}{m}\left(1 - 2zns - \frac{(s_{x} + zns^2 - s)^2}{s\left[2zm - (s_{x} + zns^2 - s)\right]}\right), \label{btm:2mch}\\
		\frac{1}{\widetilde{n}} =&\; \frac{1}{n}\left(1 - 2zns - \frac{(s_{x} + zns^2 - s)^2}{s\left[2zm - (s_{x} + zns^2 - s)\right]}\right). \label{btn:2mch}
	\end{align}
\end{subequations}
Therefore, a B\"{a}cklund transformation for equation \eqref{eq:2mch} is summarized as follows.
\begin{prop}
	Suppose that equation \eqref{eq:2mch} and the system \eqref{sxx:2mch}\eqref{st:2mch} hold, then $(\widetilde{x},\widetilde{t},\widetilde{q},\widetilde{r},\widetilde{m},\widetilde{n})$ defined by \eqref{btxt:2mch}-\eqref{btn:2mch} is a solution in parametric expressions to equation \eqref{eq:2mch}.
\end{prop}

\begin{remark}
	In the spirit of remark \ref{rem:3}, under the constraint \eqref{cons:2mch} formulas \eqref{sxx:2mch}\eqref{st:2mch} are reduced to
	\begin{subequations}
		\begin{align}
			s_{xx} =&\; \frac{s_{x}^2}{2s} + z\left(ms^2\right)_{x} + \frac{m_{x}}{m}\left(s_{x} - zms^2 -s \right) - \frac{z^2}{2}m^2 s^3 - \left(2z^2m^2 - \frac{1}{2}\right)s, \label{sxx:mch}\\
			s_{t} =&\; - \frac{\left(u_{x} + u\right)}{8z^3m^2}\left(s_{x} - zms^2 -s \right)^2 - \frac{\left(s - 2zu_{x} - 2zu\right)}{4z^3m}\left(s_{x} - zms^2 -s \right) \notag \\
			&\;  - (u_{x}^2 - u^2) s_{x} + \frac{s}{2z^2}, \label{st:mch}
		\end{align}
	\end{subequations}
	while from \eqref{btxt:2mch}-\eqref{btn:2mch} we derive
	\begin{subequations}
		\begin{align}
			\widetilde{x} =&\; x + \ln \frac{\left[2zm - (s_{x} - zms^2 - s)\right]^2}{4z^2m^2},\qquad \widetilde{t} = t, \label{btxt:mch}\\
			\widetilde{u} =&\; (u_{x} + u) \frac{\left[2zm - (s_{x} - zms^2 - s)\right]^2}{8z^2m^2} -  (u_{x} - u) \frac{2z^2m^2}{\left[2zm - (s_{x} - zms^2 - s)\right]^2} \notag \\
			&\; - \frac{\left[2zm - (s_{x} - zms^2 - s)\right]s}{4z^2m} + \frac{(s_{x} - zms^2 - s)^2}{2zs\left[2zm - (s_{x} - zms^2 - s)\right]^2}, \label{btu:mch}\\
			\frac{1}{\widetilde{m}} =&\; \frac{1}{m}\left(1 + 2zms - \frac{(s_{x} - zms^2 - s)^2}{s\left[2zm - (s_{x} - zms^2 - s)\right]}\right), \label{btm:mch}
		\end{align}
	\end{subequations}
	which could also be derived from \eqref{nxt:mch}-\eqref{nm:mch}. Formulas \eqref{btxt:mch}-\eqref{btm:mch}, together with \eqref{sxx:mch}\eqref{st:mch}, constitute a B\"{a}cklund transformation of the modified CH equation \eqref{eq:mch}. This result is equivalent to the B\"{a}cklund transformation obtained in \cite{litian22}, but different from that constructed in \cite{wangliumao20}.
\end{remark}

\section{Conclusions and discussions}\label{sec:5}
Nonlocal infinitesimal symmetries depending on eigenfunctions of linear (adjoint) spectral problems were constructed for the 2-CH equation \eqref{eq:2ch} and equation \eqref{eq:2mch}. The nonlocal infinitesimal symmetry \eqref{redinfsym:2ch} was prolonged to the enlarged system consisting of \eqref{eq:2ch}, \eqref{sp:2ch:a}\eqref{sp:2ch:b} and \eqref{p:2ch:a}\eqref{p:2ch:b}, and integrated to generate the finite symmetry transformation \eqref{nvar:2ch}, which enables us to derive a nontrivial solution and a B\"{a}cklund transformation for the 2-CH equation \eqref{eq:2ch}. Based on the nonlocal infinitesimal symmetry \eqref{redinfsym:2mch}, a finite symmetry transformation was established for the enlarged system \eqref{eq:2mch}\eqref{sp:2mch:a}\eqref{sp:2mch:b} and \eqref{px:2mch}\eqref{pt:2mch}. Consequently, a nontrivial solution and a B\"{a}cklund transformation was constructed for equation \eqref{eq:2mch}.

As explained in remark \ref{rem:2}, it seems difficulty to prolong nonlocal infinitesimal symmetries \eqref{infsym:2mch} or \eqref{redinfsym:2mch} to \eqref{eq:sqq}. It would be interesting to investigate nonlocal symmetries of \eqref{eq:sqq} via other approaches. A challenging task is to analyze complete sets of point symmetries of the enlarged system in each case, as more solutions would be yielded by symmetry reductions.

\section*{Acknowledgements}
This work was supported by the National Natural Science Foundation of China (NNSFC) (Grant Nos. 12171474, 11931017).

%
%
%

\bibliographystyle{unsrt}
\bibliography{NS2CHTRefs}

\begin{thebibliography}{10}

\bibitem{camaholm93}
R.~Camassa and D.~D. Holm.
\newblock An integrable shallow water equation with peaked solitons.
\newblock {\em Phys. Rev. Lett.}, 71(11):1661--1664, 1993.

\bibitem{misio98}
G.~Misio{\l}ek.
\newblock A shallow water equation as a geodesic flow on the {Bott-Virasoro}
  group.
\newblock {\em J. Geom. Phys.}, 24(3):203--208, 1998.

\bibitem{kour99}
S.~Kouranbaeva.
\newblock The {Camassa–Holm} equation as a geodesic flow on the
  diffeomorphism group.
\newblock {\em J. Math. Phys.}, 40(2):857--868, 1999.

\bibitem{reyes02}
E.~G. Reyes.
\newblock Geometric integrability of the {Camassa-Holm} equation.
\newblock {\em Lett. Math. Phys.}, 59:117--131, 2002.

\bibitem{reyes07}
E.~G. Reyes.
\newblock On nonlocal symmetries of some shallow water equations.
\newblock {\em J. Phys. A: Math. Theor.}, 40(17):4467–4476, 2007.

\bibitem{bies12}
P.~M. Bies, P.~G\'{o}rka, and E.~G. Reyes.
\newblock The dual modified {Korteweg-de Vries–Fokas–Qiao} equation:
  geometry and local analysis.
\newblock {\em J. Math. Phys.}, 53(7):073710 (20pp), 2012.

\bibitem{lou22}
S.~Y. Lou.
\newblock Nonlocal symmetries of nonlinear integrable systems.
\newblock In N.~Euler and D.~J. Zhang, editors, {\em Nonlinear Systems and
  Their Remarkable Mathematical Structures Volume 3 Contributions from China},
  pages 158--170, New York, 2022. Chapman and Hall/CRC.

\bibitem{reyes09}
R.~Hern\'{a}ndez-Heredero and E.~G. Reyes.
\newblock Nonlocal symmetries and a {Darboux} transformation for the
  {Camassa–Holm} equation.
\newblock {\em J. Phys. A: Math. Theor.}, 42(18):182002 (9pp), 2009.

\bibitem{litian22}
N.~H. Li and K.~Tian.
\newblock Nonlocal symmetries and {Darboux} transformations of the
  {Camassa–Holm} equation and modified {Camassa–Holm} equation revisited.
\newblock {\em J. Math. Phys.}, 63(4):041501 (8pp), 2022.

\bibitem{galas92}
F.~Galas.
\newblock New non-local symmetries with pseudopotentials.
\newblock {\em J. Phys. A: Math. Gen.}, 25(15):L981--L986, 1992.

\bibitem{consivan08}
A.~Constantin and R.~I. Ivanov.
\newblock On an integrable two-component {Camassa–Holm} shallow water system.
\newblock {\em Phys. Lett. A}, 372(48):7129–7132, 2008.

\bibitem{olvros96}
P.~J. Olver and P.~Rosenau.
\newblock {Tri-Hamiltonian} duality between solitons and solitary-wave
  solutions having compact support.
\newblock {\em Phys. Rev. E}, 53(2):1900--1906, 1996.

\bibitem{chenliuzhang06}
M.~Chen, S.~Q. Liu, and Y.~J. Zhang.
\newblock A two-component generalization of the {Camassa-Holm} equation and its
  solutions.
\newblock {\em Lett. Math. Phys.}, 75:1--15, 2006.

\bibitem{falqui06}
G.~Falqui.
\newblock On a {Camassa–Holm} type equation with two dependent variables.
\newblock {\em J. Phys. A: Math. Gen.}, 39(2):327–342, 2006.

\bibitem{wu06}
C.~Z. Wu.
\newblock On solutions of the two-component {Camassa-Holm} system.
\newblock {\em J. Math. Phys.}, 47(8):083513 (11pp), 2006.

\bibitem{holmivan11}
D.~D. Holm and R.~I. Ivanov.
\newblock Two-component {CH} system: inverse scattering, peakons and geometry.
\newblock {\em Inverse Prob.}, 27(4):045013 (19pp), 2011.

\bibitem{matsuno17}
Y.~Matsuno.
\newblock Multisoliton solutions of the two-component {Camassa–Holm} system
  and their reductions.
\newblock {\em J. Phys. A: Math. Theor.}, 50(34):345202 (28pp), 2017.

\bibitem{wangliliu20}
G.~H. Wang, N.~H. Li, and Q.~P. Liu.
\newblock Multi-soliton solutions of a two-component {Camassa–Holm} system:
  {Darboux} transformation approach.
\newblock {\em Commun. Theor. Phys.}, 72(4):045003 (6pp), 2020.

\bibitem{wangThesis21}
G.~H. Wang.
\newblock {\em Darboux transformations and {B\"{a}cklund} transformations for
  several {Camassa-Holm} type equations}.
\newblock {PhD} thesis (in chinese), China University of Mining and Technology,
  Beijing, 2021.

\bibitem{wang23}
G.~H. Wang.
\newblock Multi-soliton solutions of the two-component {Camassa–Holm}
  equation and its reductions.
\newblock {\em Theor. Math. Phys.}, 214(3):308–333, 2023.

\bibitem{tianliu13}
K.~Tian and Q.~P. Liu.
\newblock {Tri-Hamiltonian} duality between the {Wadati-Konno-Ichikawa}
  hierarchy and the {Song-Qu-Qiao} hierarchy.
\newblock {\em J. Math. Phys.}, 54(4):043513 (10pp), 2013.

\bibitem{qiao06}
Z.~J. Qiao.
\newblock A new integrable equation with cuspons and {W/M}-shape-peaks
  solitons.
\newblock {\em J. Math. Phys.}, 47(11):112701 (9pp), 2006.

\bibitem{sqq11}
J.~F. Song, C.~Z. Qu, and Z.~J. Qiao.
\newblock A new integrable two-component system with cubic nonlinearity.
\newblock {\em J. Math. Phys.}, 52(1):013503 (9pp), 2011.

\bibitem{chang16}
X.~K. Chang, X.~B. Hu, and J.~Szmigielski.
\newblock Multipeakons of a two-component modified {Camassa–Holm} equation
  and the relation with the finite {Kac-van Moerbeke} lattice.
\newblock {\em Adv. Math.}, 299:1--35, 2016.

\bibitem{zhoutianli22}
H.~Y. Zhou, K.~Tian, and N.~H. Li.
\newblock Four super integrable equations: nonlocal symmetries and
  applications.
\newblock {\em J. Phys. A: Math. Theor.}, 55(22):225207 (24pp), 2022.

\bibitem{KrasilVinog89}
I.~S. Krasil’shchik and A.~M. Vinogradov.
\newblock Nonlocal trends in the geometry of differential equations:
  symmetries, conservation laws, and {B\"{a}cklund} transformations.
\newblock {\em Acta Appl. Math.}, 15:161--209, 1989.

\bibitem{schiff96}
J.~Schiff.
\newblock Zero curvature formulations of dual hierarchies.
\newblock {\em J. Math. Phys.}, 37(4):1928--1938, 1996.

\bibitem{wangliumao20}
G.~H. Wang, Q.~P. Liu, and H.~Mao.
\newblock The modified {Camassa-Holm} equation: {B\"{a}cklund} transformation
  and nonlinear superposition formula.
\newblock {\em J. Phys. A: Math. Theor.}, 53(29):294003 (15pp), 2020.

\end{thebibliography}

\end{document}